\documentclass{PoS}

\usepackage{dsfont}

\title{Structure and properties of the vacuum of the Twisted
Eguchi-Kawai model}

\ShortTitle{Structure and properties of the vacuum of the Twisted
Eguchi-Kawai model}

\author{\speaker{H\'{e}lvio Vairinhos}
\\University of Oxford\\E-mail: \email{helvio@thphys.ox.ac.uk}}

\author{Michael Teper\\University of Oxford\\E-mail:
\email{teper@thphys.ox.ac.uk}}

\abstract{We investigate numerically the phase structure of the
Twisted Eguchi-Kawai (TEK) model in four dimensions. In the numerical
simulations of the zero-temperature TEK model (using a symmetric
twist) we observe the existence of new phases that break its $Z_N^4$
symmetry at intermediate lattice couplings and for large SU($N$)
gauge groups. This effect can be explained by the contribution of
diagonal configurations with collapsed eigenvalues, which are
particular cases of ``generalised fluxons''. We also investigate
finite-temperature versions of the TEK model using anisotropic
lattice couplings, where in particular we find van Baal fluxons
contributing at large anisotropies.}

\FullConference{The XXV International Symposium on Lattice Field
Theory\\July 30-4 August 2007\\Regensburg, Germany}

\begin{document}

\section{Introduction}

Eguchi-Kawai (EK) reduced models offer a potentially useful
alternative to the usual approach to large-$N$ extrapolations in
lattice gauge theory. In four dimensions, they are defined as SU($N$)
Wilson lattice gauge theories living on a $1^4$ lattice. In the
large-$N$ limit, these models were conjectured \cite{EK} to be
equivalent to SU($N$) Wilson lattice gauge theories living on an
infinite lattice; this equivalence is known as EK correspondence. In
practice, the absence of space-time degrees of freedom in EK models
allows the simulation of relatively large gauge groups with moderate
computational effort, which alludes to a faster convergence to the
planar limit. However, the EK correspondence doesn't always hold: in
the original EK model \cite{EK}, i.e. with periodic boundary
conditions, the EK correspondence is known to break at weak coupling.
In the EK model with twisted boundary conditions \cite{TEK}, on the
other hand, it is generally assumed that the problems that affect the
untwisted EK model are absent, and consequently that the EK
correspondence holds for all couplings. In this talk we summarise the
results of our study \cite{TeperVairinhos} of the properties of some
four-dimensional TEK models, which have strong implications for the
validity of the EK correspondence.

\section{The EK model and its vacuum}
\label{sec:EK}%
\vskip -2mm

The action of the four-dimensional EK model \cite{EK} is given by:
\begin{equation}
S_\mathrm{EK}(U) = bN \sum_{\mu\neq\nu}^4 \mathrm{Tr} \left\{
\mathds{1} - U^{}_\mu U^{}_\nu U^\dag_\mu U^\dag_\nu \right\}
\label{eq:Action:EK}%
\end{equation}
where $b=\frac{1}{g^2 N}$ is the inverse 't Hooft coupling, and
$U_\mu\in$ SU($N$) are the link variables. The EK action has a gauge
symmetry:
\begin{equation}
U_\mu \mapsto V U_\mu V^\dag, \ V \in \mathrm{SU}(N)
\label{eq:GaugeSymmetry}%
\end{equation}
and also a $Z_N^4$ symmetry:
\begin{equation}
U_\mu \mapsto z_\mu U_\mu, \ z_\mu \in Z_N
\label{eq:ZNSymmetry}%
\end{equation}
The EK correspondence \cite{EK} states that SU($N$) Wilson lattice
gauge theories living on an infinite lattice should be equivalent to
SU($N$) EK models in the large-$N$ limit, in the sense that
expectation values of corresponding operators coincide in that limit,
i.e. $\langle \mathcal{O}[U_\mu(x)] \rangle_{\rm W}
\stackrel{N\to\infty}{=} \langle \mathcal{O}[U_\mu] \rangle_{\rm EK
}$. This is a consequence of the fact that the large-$N$
Schwinger-Dyson (SD) equations of the original Wilson theory and its
EK reduced model coincide. More precisely, the SD equations have the
same form in both theories, except for contact terms that are
specific of the EK model. These contact terms consist of expectation
values of open lines; they vanish exactly in the original Wilson
theory (due to gauge symmetry), but not necessarily in the EK model
(since the gauge symmetry (\ref{eq:GaugeSymmetry}) is a similarity
transformation of the link variables). However, in the EK model the
open lines are not invariant under the $Z_N^4$ symmetry
(\ref{eq:ZNSymmetry}). So, unless the centre symmetry is
spontaneously broken, the expectation values of open lines is zero
and the EK correspondence holds nonperturbatively.

In the strong coupling regime, $b \to 0$, the Haar measure dominates
the partition function of the EK model. This measure has a repulsive
effect on the eigenvalues of the link variables, which results in an
essentially uniform distribution of eigenvalues over the unit
circle;\footnote{Recall that eigenvalues of SU($N$) matrices are pure
U(1) phases.} consequently, the traces of link variables (and other
open lines) have zero expectation value, and hence the EK
correspondence holds in the weak coupling regime.

In the weak coupling regime, $b \to \infty$, small fluctuations
around the absolute minimum of the EK action (\ref{eq:Action:EK})
dominates the partition function. Let $U_\mu = \Omega_\mu e^{-X_\mu}$
denote a perturbation (labelled by anti-Hermitian matrices $X_\mu$)
of the reduced link variables around an extremum $\Omega_\mu$ of the
EK action. The stationarity points of the action ($\delta S_{\rm
EK}=0$) are SU($N$) diagonal matrices, $\Omega_\mu={\rm diag}
\{e^{i\varphi_1}, \ldots, e^{i\varphi_N}\}$. The trace of a reduced
link variable averaged over the classical vacuum manifold is
obviously zero. However, the second variation of the EK action around
these diagonal extrema is:
\begin{equation}
\delta^2 S_{\rm EK} \propto - \sum_{\mu\neq\nu}^4 \mathrm{Tr} \left\{
\left( \Omega_\mu X_\nu \Omega^\dag_\mu - X_\nu \right) - \left(
\Omega_\nu X_\mu \Omega^\dag_\nu - X_\mu \right) \right\}^2 \geq 0
\label{eq:SecondVariation}%
\end{equation}
When the extremum is a centre element, $\Omega_\mu = z_\mu \mathds{1}
\in Z_N$, we have $\delta^2 S_{\rm EK}(z_\mu \mathds{1}) \equiv 0$,
while for general diagonal matrices we have $\delta^2 S_{\rm
EK}(\Omega_\mu)
> 0$. This means that the elements of the centre, having lower action,
are the true vacua of the quantum EK model. These vacua break the
$Z_N^4$ symmetry spontaneously, and hence the EK correspondence does
not hold in the continuum limit.

\section{The TEK model and its vacuum}
\vskip -2mm

An elegant way to avoid the spontaneous breaking of the $Z_N^4$
symmetry in the EK model is to introduce twisted boundary conditions
in the reduced $1^4$ lattice, as pioneered by Gonz\'{a}lez-Arroyo and
Okawa \cite{TEK}. The action of the twisted EK (TEK) model is given
by:
\begin{equation}
S_\mathrm{TEK}(U) = bN \sum_{\mu\neq\nu}^4 \mathrm{Tr}
\left(\mathds{1} - z_{\mu\nu} U_\mu U_\nu U^\dag_\mu U^\dag_\nu
\right)
\label{eq:Action:TEK}%
\end{equation}
where $z_{\mu\nu}=e^{i \frac{2\pi}{N} n_{\mu\nu}}$, and
$n_{\mu\nu}=-n_{\nu\mu}=L$, for all $\mu<\nu$, is the symmetric twist
tensor of Gonz\'{a}lez-Arroyo and Okawa \cite{TEK}, with $N=L^2$. The
EK correspondence states that the SU($N$) TEK model with symmetric
twist tensor has the same planar limit as the SU($N$) Wilson lattice
gauge theory living in a periodic $L^4$ lattice.

In the strong coupling regime, the Haar measure dominates the TEK
partition function. This implies, like in the EK model, that the
expectation values of the traces of open lines are zero.

In the weak coupling regime, small fluctuations around the classical
minimum of (\ref{eq:Action:TEK}) dominate the TEK partition function.
This classical minimum, known as the twist-eater, is the
configuration $U_\mu=\Gamma_\mu$ that solves the equation
$e^{i\frac{2\pi}{L}} \Gamma_\mu \Gamma_\nu \Gamma^\dag_\mu
\Gamma^\dag_\nu = \mathds{1}$; it is given, up to gauge rotations
(\ref{eq:GaugeSymmetry}) and $Z_N$ shifts (\ref{eq:ZNSymmetry}), by
$\Gamma_\mu \sim \bigoplus^L {\rm diag} \{1,e^{i\frac{2\pi}{L}},
\ldots, e^{i\frac{2\pi}{L}(L-1)}\}$. Small fluctuations around
$\Gamma_\mu$ correspond to small fluctuations around these classical
eigenvalues. The distribution of eigenvalues of a link variable is
then $Z_L$-symmetric over the unit circle, and hence its trace is
automatically zero. In sum, small fluctuations around the classical
minimum of the TEK action preserve enough of the $Z_N^4$ symmetry
(namely $Z_L^4$), for the EK correspondence to hold in the weak
coupling regime of the TEK model (unlike in the untwisted EK model).

It is generally assumed that the $Z_N^4$ symmetry of the TEK model is
intact for all couplings. However, since the arguments used for the
weak and strong coupling regimes are distinct, there is no a priori
reason why the centre symmetry should not be broken at intermediate
couplings.

\section{$Z_N^4$ symmetry breaking}
\vskip -2mm

\begin{figure}
\vskip 35mm%
\includegraphics[scale=.61,angle=0]{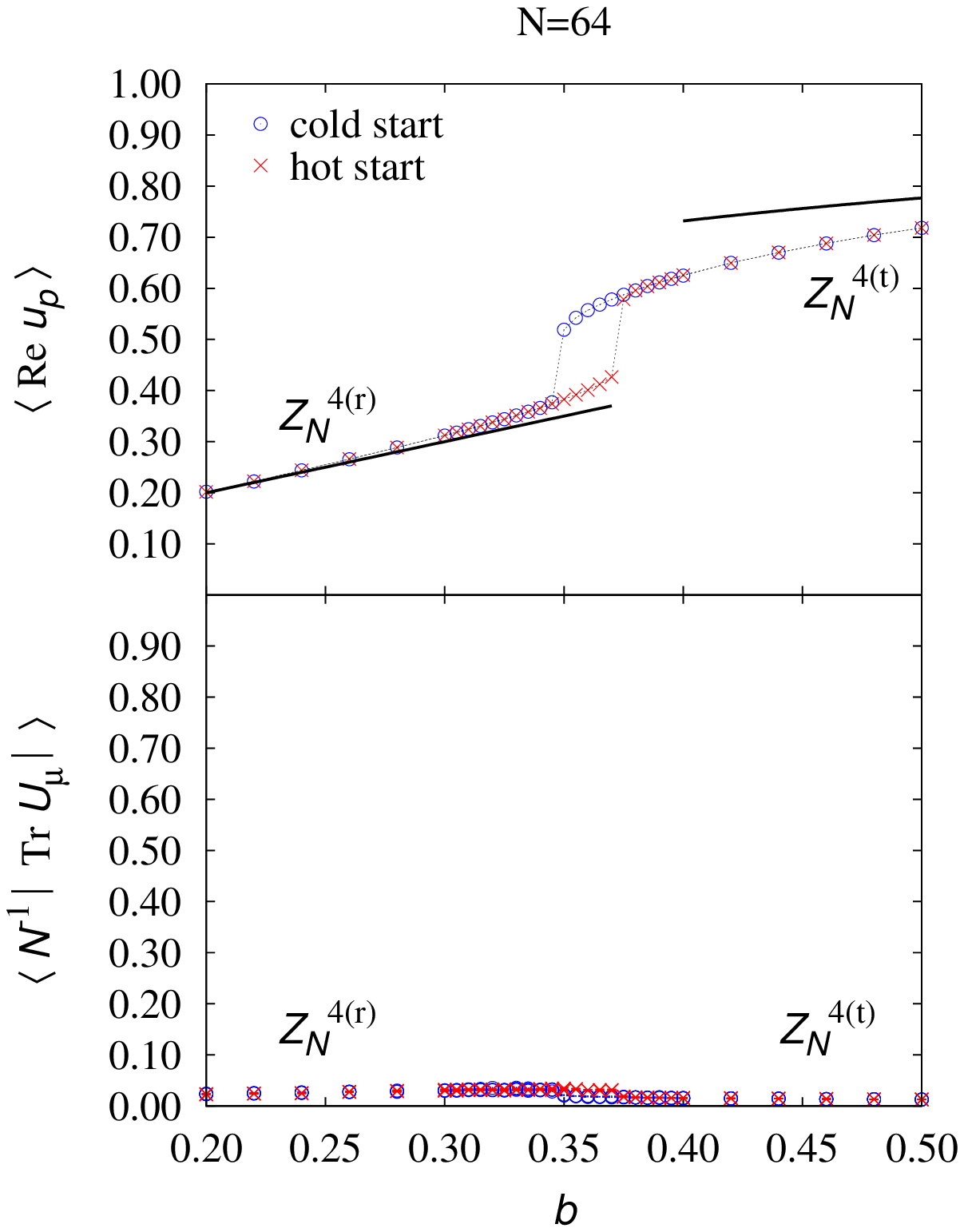}
\includegraphics[scale=.61,angle=0]{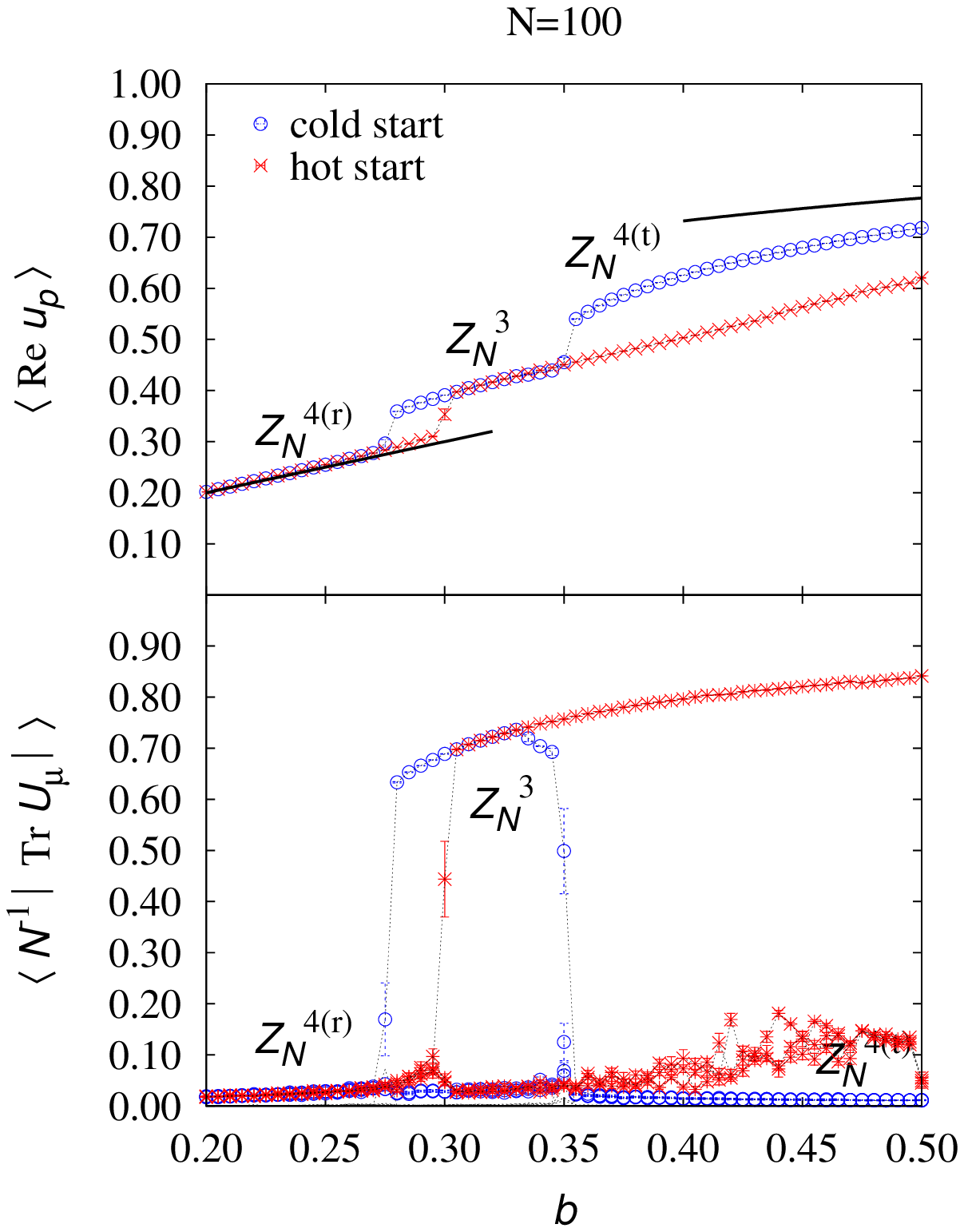}
\caption{Expectation values of the plaquette, $u_p$, (top) and traced
link variables, ${\rm Tr}~ U_\mu$, (bottom) in the SU($N$) TEK models
for $N=64$ (left) and $N=100$ (right). $Z_N^{k}$ denotes a phase with
$k$ preserved $Z_N$ symmetries; $Z_N^{4(r)}$ and $Z_N^{4(t)}$ denote
phases that are $Z_N^4$-symmetric due to the randomisation from the
Haar measure $(r)$ or due to fluctuations around the twist-eater
vacuum $(t)$, respectively.}%
\label{fig:ZNBreaking}%
\end{figure}

We performed numerical simulations of the SU($N$) TEK model with
symmetric twist for a wide range of gauge groups, namely $25 \leq N
\leq 256$. For $N \leq 81$, the TEK model has properties similar to
the ones of the original Wilson theory: it has a strong first-order
bulk transition at $b \approx 0.36$, and agrees with the strong- and
weak-coupling expansions of the plaquette (left-top graph in
Fig.\ref{fig:ZNBreaking}). In sum, the EK correspondence seems to
hold without restrictions, which is also suggested by the fact that
the trace of reduced link variables is zero for all couplings
(left-bottom graph in Fig.\ref{fig:ZNBreaking}). However, for $N \geq
100$, the $Z_N^4$ symmetry of the TEK model is spontaneously broken
at intermediate couplings; this is manifest by at least one of the
link variables acquiring a non-zero trace (right-bottom graph in
Fig.\ref{fig:ZNBreaking}). In hot start simulations, we also observe
that once the reduced lattice falls into a $Z_N$-breaking vacuum, it
remains there for the whole simulation, even for very large values of
the coupling ($b \approx 20$). This metastability suggests the
existence of stable extrema of the TEK action that do not preserve
its $Z_N^4$ symmetry.

\begin{figure}
\vskip 30mm%
\includegraphics[scale=.49,angle=0]{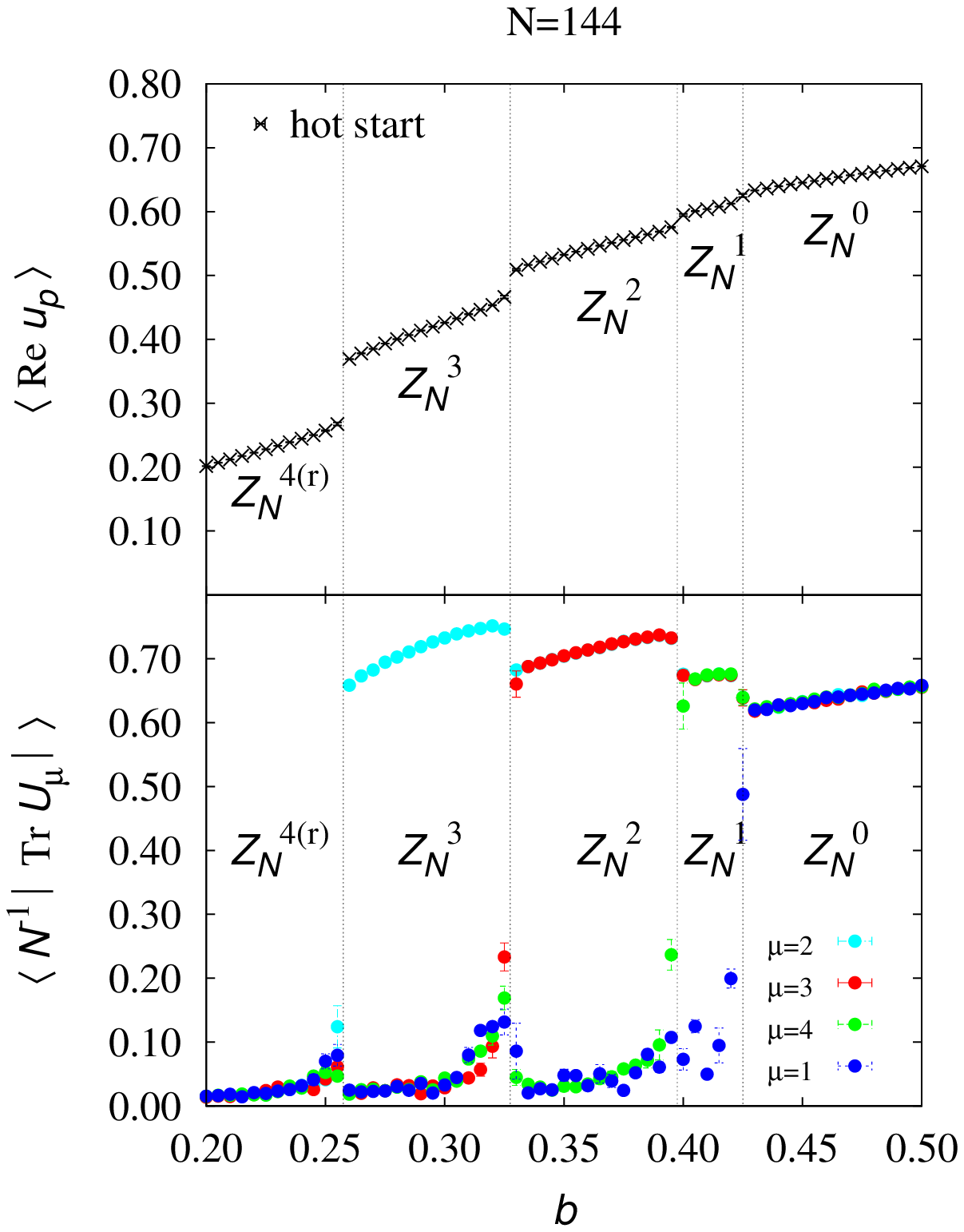}
\hskip 9mm%
\includegraphics[scale=.48,angle=0]{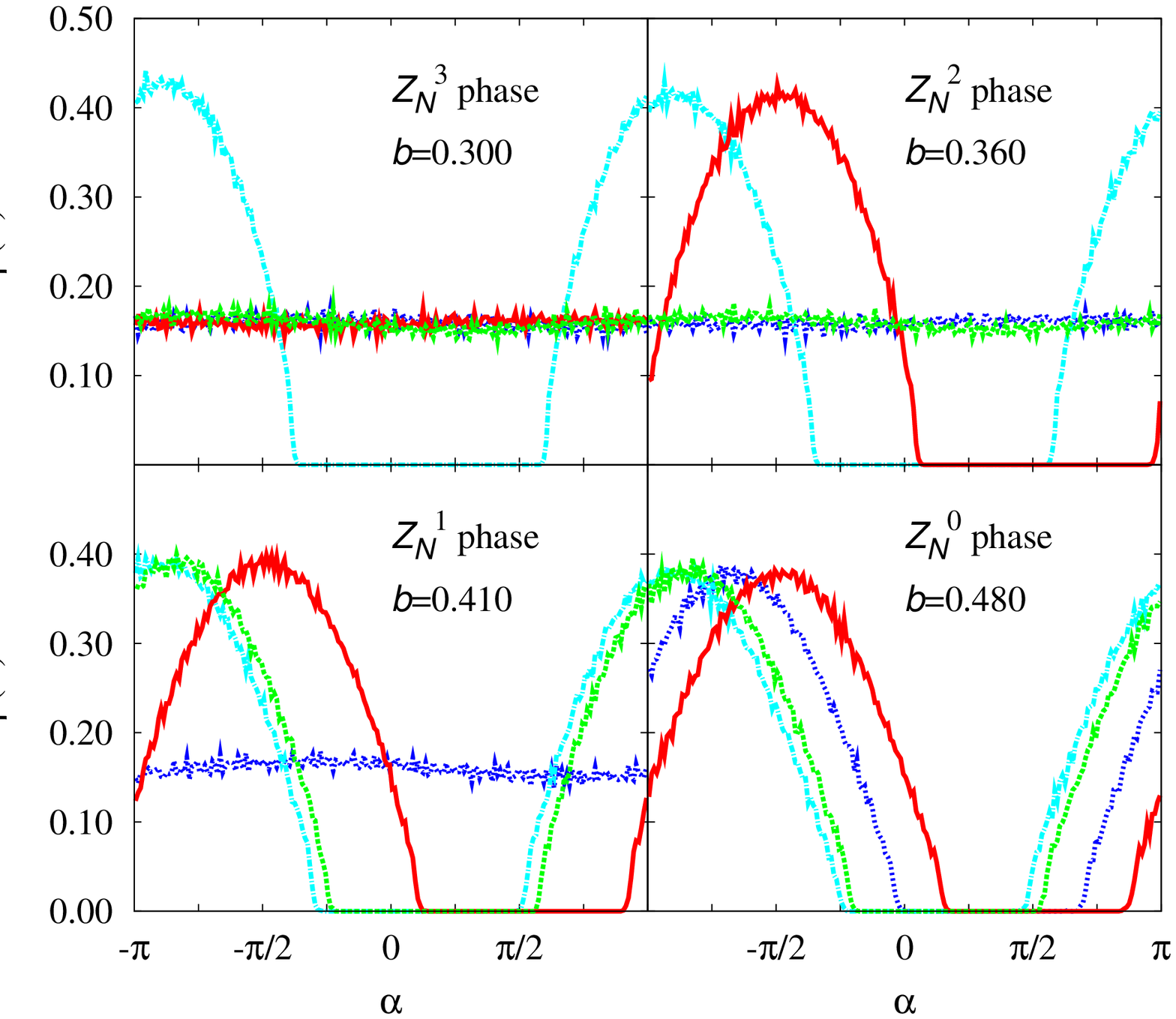}
\caption{Left: Expectation values of the plaquette, $u_p$, (top) and
traced link variables, ${\rm Tr}~ U_\mu$, (bottom) in the SU($N$) TEK
models for $N=144$. Right: Eigenvalue distributions, $\rho(\alpha)$,
of the link variables in the different $Z_N^k$-symmetric phases of
the SU(144) TEK model ($k=0,1,2,3$).}
\label{fig:EigTransition}%
\end{figure}

We also observe that the four independent $Z_N$ symmetries of the TEK
model break independently in a cascading way: $Z_N^4 \to Z_N^3 \to
Z_N^2 \to Z_N^1 \to Z_N^0$ (Fig.\ref{fig:EigTransition}). These
transitions are similar to the ones observed on a continuum torus by
Narayanan and Neuberger \cite{NN}. However, they are not the physical
transitions of the original Wilson theory, but instead the
transitions associated with the $1^4$ lattice (where the link
variables play the role of Polyakov loops). Across each one of these
transition, the eigenvalues of one of the link variables collapse
onto an element of $Z_N$, which consequently leads to a non-zero
expectation value for its trace. This suggests that the stable
relative extrema of the TEK action that break the $Z_N^4$ symmetry
are centre configurations.

The existence of $Z_N$-breaking phases in the TEK model has been
confirmed independently by Ishikawa and Okawa \cite{IshikawaOkawa},
and has also been also observed in the context of discrete
noncommutative gauge theories \cite{Bietenholz:LAT07}.

\section{Interpretation}
\vskip -2mm

A class of stable extrema of the TEK action (\ref{eq:Action:TEK}) are
the unitary solutions $\Omega_\mu$ of the equation
$e^{i\frac{2\pi}{L}} \Omega_\mu \Omega_\nu \Omega^\dag_\mu
\Omega^\dag_\nu = e^{i\frac{2\pi}{N}k_{\mu\nu}} \mathds{1}$, where
$\cos\left(\frac{2\pi}{N}k_{\mu\nu}\right) \geq 0$ for all $\mu<\nu$
(stability condition). These solutions are called
`fluxons'\footnote{Normally, the term `fluxon' is only used when the
solution for a given $k_{\mu\nu}$ gives a non-zero contribution to
the partition function in the $N\to\infty$ limit. Here we abuse the
language and call `fluxon' any solution of this equation.}
\cite{Baal:Surviving}. Solutions of the same `fluxon level', $k =
\frac{1}{2} \sum_{\mu<\nu}^4 k_{\mu\nu}^2 \in \mathds{Z}$, are
gauge-equivalent and have the same classical action, $S_{\rm TEK}
\propto k$. The classical minimum of the TEK action -- the
twist-eater -- is the solution for the special case $k_{\mu\nu}
\equiv 0$.

If we consider the case $k_{\mu\nu}=L$, then the stationarity
equation is $\Omega_\mu \Omega_\nu \Omega^\dag_\mu \Omega^\dag_\nu =
\mathds{1}$, whose solution manifold is the set of all diagonal
matrices. This is a stable minimum of the TEK action. As in the case
of the untwisted EK model (Section \ref{sec:EK}), fluctuations around
this extremum result in the centre configurations ($z_\mu\mathds{1}$)
having a lower action than the general diagonal matrix. Therefore,
the $Z_N^4$ symmetry is also spontaneously broken in these extrema.
The numerical data suggests that at intermediate couplings the TEK
model is dominated by diagonal matrices, even though these extrema do
not survive the large-$N$ limit (because $S_{\rm TEK} \propto N$).
However, at intermediate couplings and finite $N$, the magnitude of
the fluctuations $X_\mu$ around the twist-eater may be large enough
to overcome the barriers between extrema and eventually tunnel to a
centre configuration. While the contribution to the TEK action from
fluctuations around the twist-eater configuration is quadratic, it is
of fourth-order around centre configurations:
$$
\left\{
  \begin{array}{ll}
    S_{\rm TEK}(\Gamma_\mu) \approx - bN \sum_{\mu<\nu}^4
{\rm Tr}\ (\Gamma^\dag_\mu X_\nu \Gamma_\mu - X_\nu - \Gamma^\dag_\nu
X_\mu \Gamma_\nu + X_\mu)^2 + \cdots \equiv O(X^2) + O(X^3) \\
    S_{\rm TEK}(z_\mu\mathds{1}) \approx 24 bN^2
\sin^2(\frac{\pi}{\sqrt{N}}) - bN\sum_{\mu<\nu}^4 z_{\mu\nu} {\rm
Tr}\ ( [X_\mu,X_\nu]^2 ) + \cdots \equiv O(X^0) + O(X^4)
  \end{array}
\right.
$$
We may imagine that, at some critical coupling, the action for
fluctuations around the twist-eater may become larger than the action
for fluctuations around a centre configuration, inducing the
tunnelling between the two extrema and hence breaking the $Z_N^4$
symmetry spontaneously.

We also observe that the critical couplings of the $Z_N$-breaking
transitions (Table \ref{table:TransitionsCold}) show a dependence
with $N$ that points to a widening of the $Z_N$-broken region for
increasing $N$. In addition, we have not been able to detect any
evidence for the expected confining/deconfining transition (or any
other transitions \cite{NN}) at weak couplings.
\begin{table}
\begin{center}
\begin{footnotesize}
\begin{tabular}{lllllllllllll}\hline
$N$ &&&&&$\longleftarrow$&&&& \\\hline
100 & [$Z_N^{4(r)}$] & 0.275(5)  & [$Z_N^{3}$] &          &             & 0.350(5) &             &           & [$Z_N^{4(t)}$] \\
121 & [$Z_N^{4(r)}$] & 0.250(5)  & [$Z_N^{3}$] &          & 0.325(5)    &          & [$Z_N^{0}$] & 0.360(5)  & [$Z_N^{4(t)}$] \\
144 & [$Z_N^{4(r)}$] & 0.235(5)  & [$Z_N^{3}$] & 0.275(5) & [$Z_N^{2}$] & 0.325(5) & [$Z_N^{0}$] & 0.370(5)  & [$Z_N^{4(t)}$] \\
169 & [$Z_N^{4(r)}$] & 0.235(15) & [$Z_N^{3}$] & 0.25(3)  & [$Z_N^{2}$] & 0.355(15)& [$Z_N^{0}$] & 0.37(3)   & [$Z_N^{4(t)}$] \\
196 & [$Z_N^{4(r)}$] && 0.235(15) &          & [$Z_N^{2}$] & 0.28(3)  & [$Z_N^{0}$] & 0.385(15) & [$Z_N^{4(t)}$] \\
225 & [$Z_N^{4(r)}$] && 0.22(3)   &          & [$Z_N^{2}$] & 0.28(3)  & [$Z_N^{0}$] & 0.415(15) & [$Z_N^{4(t)}$] \\
256 & [$Z_N^{4(r)}$] & 0.205(15) & [$Z_N^{3}$] & 0.235(15)& [$Z_N^{2}$] & 0.28(3)  & [$Z_N^{0}$] & 0.43(3)   & [$Z_N^{4(t)}$] \\
    \hline
\end{tabular}
\end{footnotesize}
\caption{Critical values of $b$ associated with the
breaking/restoration of one (or more) $Z_N$ symmetries of the TEK
model, for cold start simulations. [$Z_N^k$] refers to the phase with
$k$ unbroken directions; the numerical values denote the critical
couplings; a numerical value spanning multiple columns is the
critical coupling associated with the simultaneous breaking of more
than one $Z_N$ symmetry. \label{table:TransitionsCold}}
\end{center}
\end{table}

\section{Anisotropic TEK model and fluxon vacua}
\vskip -2mm

\begin{figure}
\vskip -5mm
\hskip 14mm%
\includegraphics[scale=.60,angle=0]{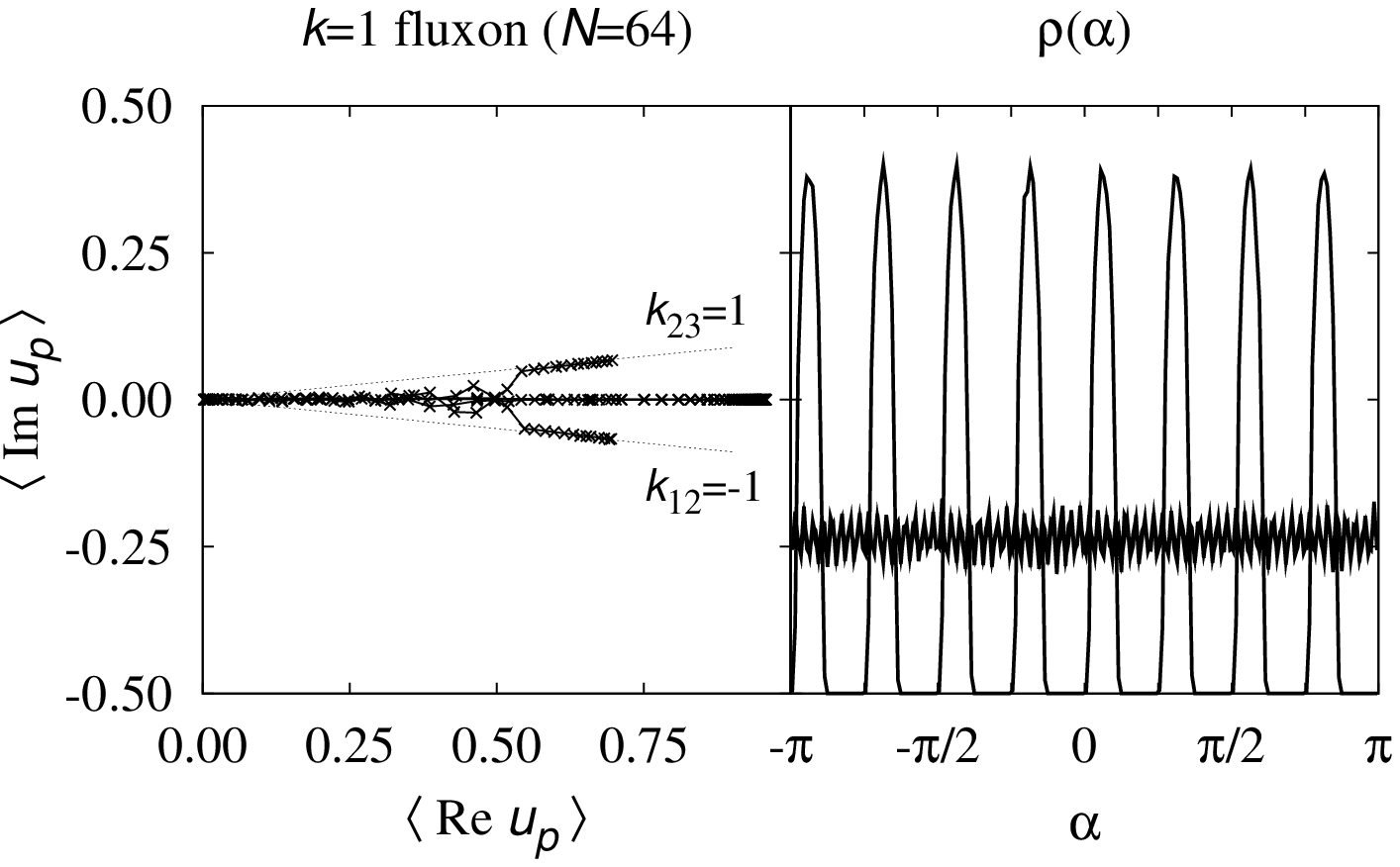}
\hskip 14mm%
\includegraphics[scale=.60,angle=0]{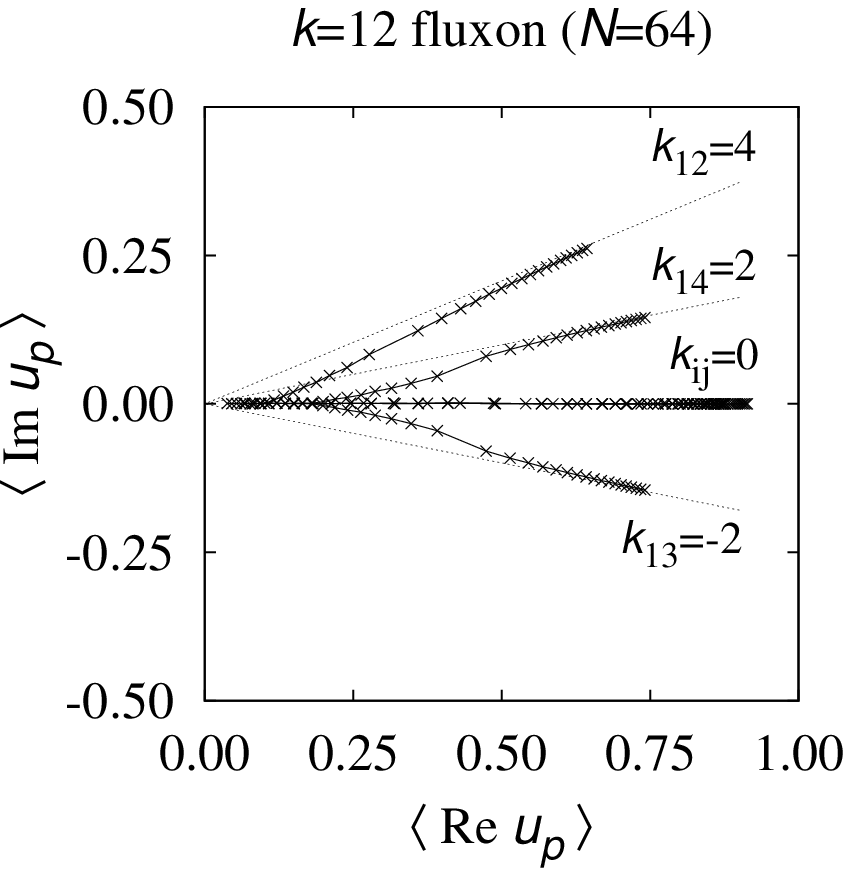}
\caption{Traces of plaquettes, $u_p$, and eigenvalue densities of
link variables, $\rho(\alpha)$, in the SU(64) anisotropic TEK model,
for different choices of $\xi_i$: $\xi_i=0.15$ (left) and
$\xi_i=(0.50,0.75,1.00)$ (right).}%
\label{fig:Fluxon}%
\end{figure}

We also simulated anisotropic TEK models, mainly to look for evidence
of the large-$N$ confining/deconfining transition at finite
temperature. The action of the most general anisotropic TEK model is
given by:
\begin{equation}
S_\mathrm{TEK}(\xi_i,U) = bN \sum_{\mu\neq\nu}^4 \xi_{\mu\nu}
\mathrm{Tr} \left(\mathds{1} - z_{\mu\nu} U_\mu U_\nu U^\dag_\mu
U^\dag_\nu \right)
\label{eq:Action:TEKani}%
\end{equation}
where $\xi_{\mu\nu}= \frac{\xi_\mu \xi_\nu}{\xi_\kappa \xi_\lambda}$
($\kappa\neq\lambda$, and $\kappa,\lambda\neq\mu,\nu$), and $\xi_\mu
= \frac{a_4}{a_\mu}$ ($a_\mu$ is the lattice spacing in the $\mu$
direction). For $N \geq 81$, we observe that the $Z_N^4$ symmetry is
broken by centre configurations. For $N \leq 64$, however, different
phases appear at intermediate coupling. By analysing the values of
the plaquettes and eigenvalue structure of the link variables in
these new extrema (Fig.\ref{fig:Fluxon}), we conclude that they are
fluxon configurations \cite{Baal:Surviving}. We observe intermediate
stable configurations belonging to several different fluxon levels,
depending on the choice of $\xi_i$; in particular, for $\xi_i=\xi$,
we observe van Baal fluxons \cite{Baal:Surviving} for the first time,
namely fluxons of the level $k=1$ (Fig.\ref{fig:Fluxon}); for
different cases, the fluxons are more general. We again found no
evidence for the confining/deconfining transition (or any other
transitions \cite{NN}) in the anisotropic TEK model.

\section{Conclusions}
\vskip -2mm

We showed that for sufficiently large $N$, the $Z_N^4$ symmetry of
SU($N$) TEK models is spontaneously broken at intermediate couplings,
where centre configurations dominate. These $Z_N$-breaking phases
appear to be extending further into weak coupling as $N$ increases.
Even though we cannot yet establish a clear trend for this
$N$-dependence, it suggests that the range of couplings of the
$Z_N$-broken phases may continue to grow and eventually dominate the
whole phase diagram of the TEK model in the $N\to\infty$ limit, thus
invalidating the EK correspondence for this model, or at least making
it impractical for the study of the physics of large-$N$ gauge
theories.

We also observed several fluxon vacua dominating at intermediate
couplings in anisotropic TEK models. This shows that the phase
diagram of the TEK model is much richer than previously thought.
However, no evidence of any transition to a physical phase was
observed, which could mean that they are absent or inaccessible.

\section*{Acknowledgements}
\vskip -2mm

Our lattice simulations were carried out on PPARC and EPSRC funded
computers in Oxford Theoretical Physics. HV is supported by FCT
(Portugal) under the grant SFRH/BD/12923/2003.

\end{document}